\documentclass{aastex}
\usepackage{spr-astr-addons}

\begin{document}


\title{Measuring the precise photometric period of the intermediate polar 1RXS~J230645.0+550816 (1SWXRT~J230642.7+550817) based on extensive photometry}
\shorttitle{The photometric period of IP 1RXS~J230645.0+550816}
\shortauthors{Kozhevnikov}

\author{V. P. Kozhevnikov}
\affil{Astronomical Observatory, Ural Federal University, Lenin Av. 51, Ekaterinburg 
620083, Russia e-mail: valery.kozhevnikov@urfu.ru}


\begin{abstract} 
Recently, \citeauthor{halpern18} discovered an oscillation with a period of 464~s in the cataclysmic variable 1RXS~J230645.0+550816. I conducted extensive photometric observations of this object to clarify the coherence of this oscillation and to measure the oscillation period with high precision. Observations were obtained over 22 nights in 2018 and 2019. The total duration of observations was 110~h. The oscillation was revealed in each long night of observations and was coherent throughout all my observations covering 15 months. Due to the large coverage of observations, I determined the oscillation period with high precision, which was $464.45600\pm0.00010$~s. The oscillation semi-amplitude was large and showed changes from $41.8\pm1.5$~mmag in 2018 to $48.6\pm1.8$~mmag in 2019. The oscillation pulse profile was symmetrical with a noticeably wider minimum compared to the maximum and showed no noticeable changes during 2018 and 2019. The high precision of the oscillation period allowed me to derive an oscillation ephemeris with a long validity of 70 years. This ephemeris can be used for future studies of oscillation period changes. Although short-period X-ray oscillations have not yet been detected in 1RXS~J230645.0+550816, the intermediate polar nature of this object is very probable due to the high degree of coherence of the 464-s oscillation.

\end{abstract} 

\keywords{Stars: individual, \\ 1RXS~J230645.0+550816; Novae, Cataclysmic variables; Stars: oscillations}
\section{Introduction}

Intermediate polars (IPs) are a subclass of cataclysmic variables (CVs) in which a magnetic white dwarf accretes material from a late-type companion filling its Roche lobe. The white dwarf spins asynchronously with the orbital motion, and in most cases accretion occurs through a truncated accretion disc. These reasons lead to many phenomena of periodic variability. Because the axis of the magnetic dipole of the white dwarf is tilted to its spin axis, the main oscillation occurs with the spin period the white dwarf. This oscillation is generated both in optical and in X-rays. The reprocessing of X-rays in the secondary star or asymmetric parts of the disc creates an oscillation with a beat period, $1/P_{\rm beat} = \omega-\Omega$, where $\omega=1/P_{\rm spin}$ and  $\Omega=1/P_{\rm orb}$. This is the lower orbital sideband of the spin frequency. The secondary star and disc cannot efficiently reflect X-rays with a beat period \citep[e.g.,][]{norton92}. However, a strong X-ray oscillation with the frequency $\omega - \Omega$ can be generated due to disc overflow accretion \citep{hellier14, hellier91, norton92}. In addition, due to the amplitude modulation caused by geometric effects, oscillations with the frequencies $\omega-2\Omega$, $\omega+\Omega$, $\omega+2\Omega$ can be observed. But these signals cannot be as strong as $\omega$ and $\omega - \Omega$ \citep{warner86}. Reviews of IPs are provided in \citet{patterson94, warner95, hellier96, hellier01, hellier14}. Reviews of X-ray properties of IPs are presented in \citet{kuulkers06, mukai17}.  The most recent data about IPs is provided in \citet{demartino20}.

Coherent X-ray oscillations with periods significantly shorter than the orbital period are the defining characteristic of IPs \citep{mukai05}. This gives full confidence in the classification as an IP. However, the optical short-period oscillation of high stability and the presence of one or more orbital sidebands, even without any X-ray detection at all, are a valid argument for inclusion in the IP list \citep{kuulkers06}. A simple and very effective way to distinguish this stable oscillation from intermittent quasi-periodic variations is to check whether the same period is detected in multiple data sets \citep{mukai17}. This also means that the oscillation is detectable with observations of a comparable quality.

At large time scales, spin periods in IPs can undergo small variations due to the interaction of braking torques and accretion torques \\ (e.g., \citealt{warner91}). Long-term tracking of the spin period allows us to check the spin equilibrium either by analysing the observed minus calculated (O--C) times of the spin pulse maxima obtained using a precise oscillation ephemeris or by direct measurements of the spin period \citep{patterson20}. Spin equilibrium can be proved due to alternating spin-up and spin-down, which occur due to accretion rate changes \citep{patterson94}. The spin equilibrium check is important because many theoretical works assume that IPs are in spin equilibrium \citep[e.g.,][]{norton04}. Observations of continuous spin-up or spin-down are also important because such measurements allow us to understand the angular momentum flows within the binary star \citep{king99}.

The source 1SWXRT~J230642.7+550817 is identified with the source 1RXS~J230645.0+550816 (hereafter J2306) and was classified as a CV in \citet{landi17}. \citet{halpern18} performed photometry of the identified star and discovered an oscillation with a period of $464.452\pm0.004$~s. Based on this oscillation, \citeauthor{halpern18} claimed that J2306 is a new IP. However, the photometric observations performed by \citeauthor{halpern18} were short and consisted of only 5 nights with a total duration of 12~hr. In addition, \citeauthor{halpern18} performed spectroscopic observations, but the coverage was not sufficient to determine a firm orbital period. The most likely period was near 0.136~d, but other daily aliases and some much longer periods remained possible.

I performed a tentative photometric observation of J2306. Despite the faintness of J2306 (17~mag), this observation demonstrated that the 464-s oscillation discovered by \citeauthor{halpern18} can be easily detected using my observation technique. The observations were continued to clarify the oscillation coherence, to determine the oscillation period with high precision and to derive the oscillation ephemeris with a long validity. In this paper, I present the results of all my observations obtained over 22 nights with a total duration of 110~hr, which cover 15 months.

\section{Observations} 

In observations of CVs I use a multi-channel pulse-counting photometer with photomultipliers, which allows me to perform continuous measurements of the brightness of two stars and the sky background. The design of the photometer is described in \\ \citet{kozhevnikoviz}. Such photometers are especially suitable for studies of rapid stellar variability because they do not have restrictions on time resolution. In contrast with CCD systems, most photoelectric photometers enable to observe relatively bright stars that are visible to the eye. Indeed, for a long time I could not observe stars fainter than 15 mag, which are invisible in my 70-cm Cassegain telescope equipped with the multichannel photometer. A few years ago, however, I learned to observe much fainter stars. A faint star that is invisible to the eye can be placed in the diaphragm of the photometer using the coordinates of the faint star and the coordinates of the nearest reference star as well as the telescope's step motors. The comparison star should be visible to the eye. To maintain accurate centring of stars in the photometer diaphragms, I use a CCD guiding system. Using this method, I can observe very faint stars up to 20~mag. Of course, for stars fainter than 18~mag, photon noise (Poisson fluctuations of counts) is significant and allows me to detect only deep eclipses in CVs \citep[e.g.,][]{kozhevnikov18}. In addition, the centring of stars fainter than 18~mag is difficult to check, whereas the centring of brighter stars can be checked and corrected using the telescope's step motors and star counts.

I conducted photometric observations of J2306 at Kourovka Observatory, Ural Federal University, which is located 80 km from Yekaterinburg. These observations were performed in 2018 and 2019 over 22 nights with a total duration of 110~h. The data were obtained in white light (approximately 3000--8000~\AA) with a time resolution of 8~s. Although this time resolution seems excessively high for analysing the 464-s oscillation discovered by \citet{halpern18}, it allows me to accurately fill in the gaps between individual observational nights. For J2306 and the comparison star, I used 16-arcsec diaphragms. To measure the sky background, I used a diaphragm of 30~arcsec. This large diaphragm diminishes the photon noise from the sky background. The comparison star is USNO-A2.0 1425-14616660. It has $B=14.2$~mag and $B-R=0.9$~mag. The colour index of J2306 is similar to the colour index of this star.  According to the USNO-A2.0 catalogue, J2306 has $B=17.3$~mag and $B-R=0.7$~mag.  The similarity of the colour indexes of these two stars diminishes the effect of differential extinction. The differential magnitudes were determined by taking into account differences in light sensitivity between the photometer cannels and differences in sky background counts caused by differences in the size of the diaphragms.

The observation journal is shown in Table~\ref{journal}. This table gives BJD$_{\rm TDB}$, which is the  Barycentric Julian Date in the Barycentric Dynamical Time (TDB) standard. BJD$_{\rm TDB}$ is preferred because this time is uniform. BJD$_{\rm TDB}$ is easily to calculate using the online calculator (http://astroutils.astronomy.ohio-state.edu/time/) \citep{eastman10}. To verify these calculations, I also calculated BJD$_{\rm UTC}$ using the BARYCEN routine in the ‘aitlib’ IDL library of the University of T\"{u}bingen (http://astro.uni-tuebingen.de/software/idl/aitlib/). \\ During my observations, the difference between \\  BJD$_{\rm TDB}$ and BJD$_{\rm UTC}$ was constant. BJD$_{\rm TDB}$ exceeded BJD$_{\rm UTC}$ by  69~s.

\begin{table}[t]
{\small 
\caption{Journal of the observations.}
\label{journal}
\begin{tabular}{@{}l c c}
\hline
\noalign{\smallskip}
Date  &  BJD$_{\rm TDB}$ start & length \\
(UT) & (-2458000) & (h) \\
\hline
2018 Oct. 9   & 401.410544 &  2.1  \\
2018 Nov. 8   & 431.099818 & 9.6  \\
2018 Nov. 9   & 432.083123 & 9.4   \\
2018 Nov. 10   & 433.081504 & 7.3  \\
2018 Nov. 11   & 434.188075 & 9.0  \\
2018 Nov. 13   & 436.289069 & 2.2  \\
2018 Dec. 7   & 460.224155 & 3.4  \\
2018 Dec. 9    & 462.248180 & 2.6 \\
2018 Dec. 15    & 468.431424 & 2.1 \\
2019 Sep. 8    & 735.268509 & 4.6  \\
2019 Sep. 22   & 749.180256 & 5.0  \\
2019 Sep. 30    & 757.157175 & 4.3  \\
2019 Oct. 4    & 761.215582 & 2.5 \\
2019 Oct. 5    & 762.181466 & 4.0 \\
2019 Oct. 6    & 763.270478 & 1.8 \\
2019 Oct. 7    & 764.277577 & 5.9 \\
2019 Nov. 19   & 807.078200 & 4.5 \\
2019 Nov. 20    & 808.072637 & 4.9 \\
2019 Nov. 26    & 814.339239 & 3.5 \\
2019 Nov. 29    & 817.082099 & 8.3 \\
2019 Dec. 29    & 847.064840 & 10.5 \\
2019 Dec. 30    & 848.066988 & 2.4 \\
\hline
\end{tabular} }
\end{table}


\begin{figure}[t]
\includegraphics[width=84mm]{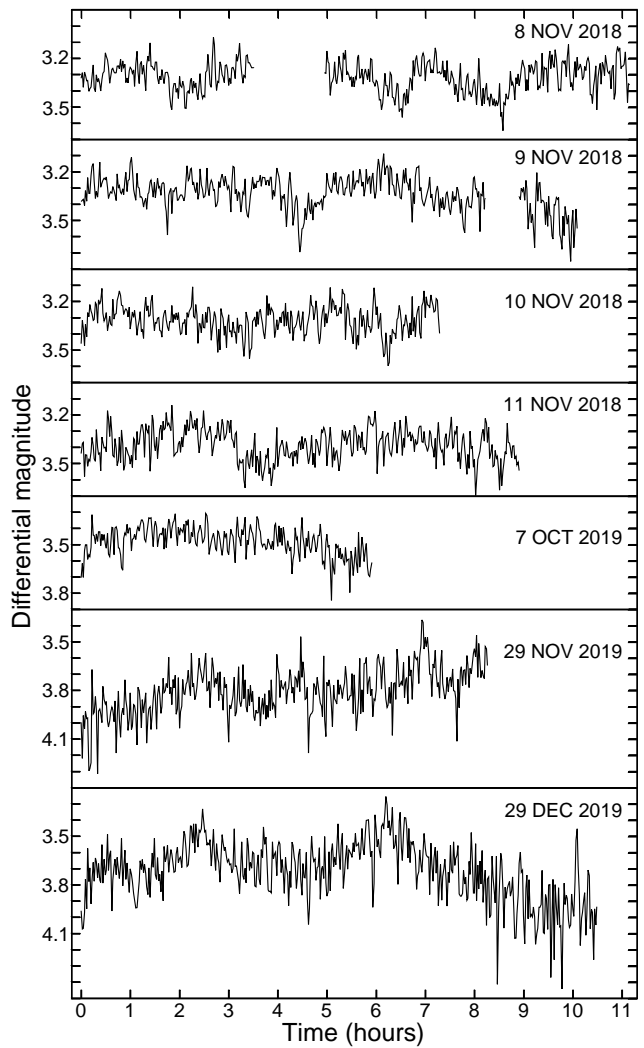}
\caption{Seven longest differential light curves of J2306. Despite the large photon noise and flickering, the short-period oscillation discovered by \citeauthor{halpern18} is noticeable in some sections of the light curves (e.g., near the end of November 11, 2018 and in the middle of October 7, 2019).}
\label{fig1}
\end{figure}

The differential light curves obtained with a time resolution of 8~s have significant photon noise due to the low brightness of J2306. To improve the visibility of the 464-s oscillation discovered by \citet{halpern18} directly in the light curves, I averaged the magnitudes over 80-s time intervals. Fig.~\ref{fig1} shows the 7 longest differential light curves of J2306 with a time resolution of 80~s. According to the counts of J2306, the comparison star and the sky background, the photon noise of these light curves (rms) is 0.04--0.07~mag. The semi-amplitude of the 464-s oscillation discovered by \citeauthor{halpern18} is comparable with this photon noise (see their Fig.~17). Another reason that prevents the direct visibility of this oscillation is the flickering seen in the light curves. From Fig.~\ref{fig1}, I estimate that the peak-to-peak flickering amplitude in J2306 is 0.3--0.4~mag. Thus, the photon noise and flickering mask the direct visibility of the 464-s oscillation discovered by \citeauthor{halpern18} Nevertheless, some sections of the light curves (Fig.~\ref{fig1}) yet show this oscillation.

\section{Analysis and results}

My multichannel photometer allows me to obtain differential magnitudes with strictly constant time intervals between them. Data having regular time sampling can be best analysed using methods based on the fast Fourier transform (FFT) \citep{schwarzenberg12}. Indeed, using the FFT, calculations are extremely fast and allows me to analyse the combined time series obtained with high time resolution, which cover time intervals reaching a year and more. In addition, because of its basic functions, Fourier analysis is preferred for sinusoidal and quasi-sinusoidal signals \citep{schwarzenberg98}. Therefore, it is most suitable for IPs, which are characterized by short-period oscillations of a smooth quasi-sinusoidal shape.

Fourier analysis is performed when the time series have a mean value equal to zero. However, simply subtracting the nightly average from the light curve is not sufficient, because the light curve may contain variations with periods exceeding its length. If these variations are not excluded, then the light curve may contain large jumps at the beginning and end. These jumps have a wide frequency content, which increases the noise level in the power spectrum. \citet{bendat86} recommended eliminating such variations by subtracting a first-, second- or third-order polynomial fitted to the time series. Individual time series that have a mean value equal to zero can be included in the combined time series with gaps between them in accordance with the observation time. Obviously, it is best to fill in the gaps with zeros because zeros do not contribute to the Fourier transform and do not introduce additional data jumps. In the subsequent analysis, I use individual light curves with a mean value equal to zero obtained by subtracting a first- or second-order polynomial fitted to the light curve.

The classical power spectrum is defined as the sum of squares of the real and imaginary parts of the Fourier transform multiplied by $2/N$, where $N$ is the number of points in the time series \citep[e.g.,][]{bendat86}. Instead, when calculating the power spectrum, I multiply this sum by $2/N^2$. The power spectrum calculated in this way has a simple meaning.  The height of the peak of the coherent oscillation is $A^2/2$, where $A$ is the semi-amplitude of the oscillation, and the sum of all points of the power spectrum is equal to the variance of the time series. This is a consequence of the Parseval's theorem \citep[e.g.,][]{kozhevnikoviz}. It is obvious that the noise level in this power spectrum decreases as the number of points in the time series increases. This means that the ability to detect a coherent oscillation hidden in noise also increases with the number of points. For ease of use, the power spectrum can be converted to an amplitude spectrum that has a less extended vertical axis and shows the semi-amplitude directly instead of its square divided by 2. 

Usually, to use the FFT, the time series must contain a number of points equal to a power of 2, $n=2^p$, where $p$ is an integer. Then the number of frequency components in the power spectrum is $n/2$. In this case, the maximum frequency resolution is provided, $\Delta f=1/T$, where $T$ is the length of the time series \citep[e.g.,][]{bendat86}, and this power spectrum covers all frequencies from the lowest frequency $1/T$ to the Nyquist limit $1/2t$, where $t$ is the time resolution. 

\begin{figure}[t]
\includegraphics[width=84mm]{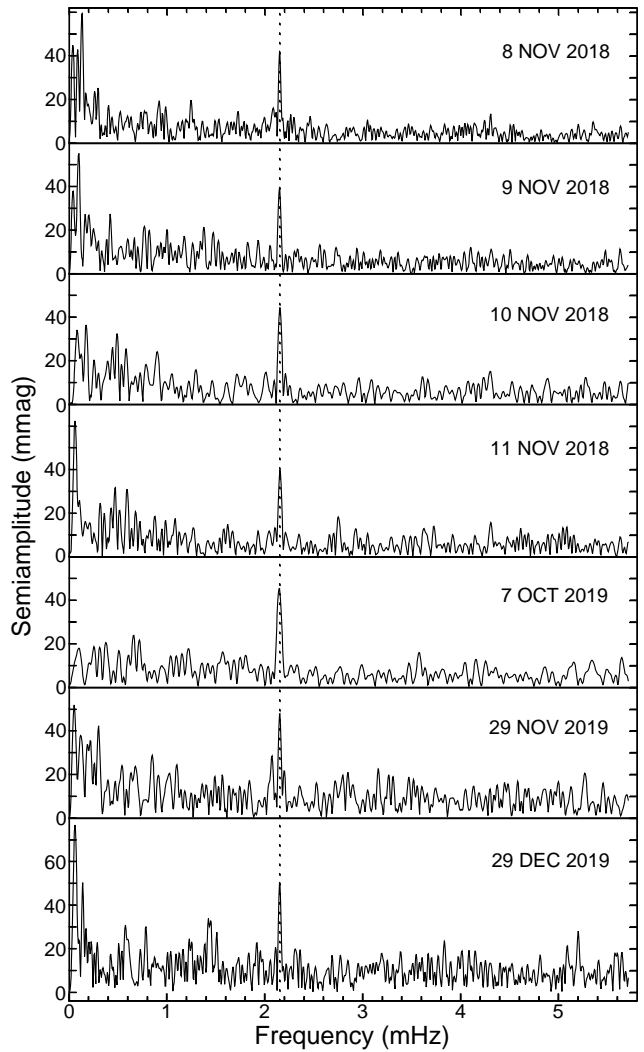}
\caption{Low-frequency parts of the amplitude spectra of the seven longest differential light curves of J2306. All of them show persistent peaks with a period of 464~s.}
\label{fig2}
\end{figure}

The precision of determining the oscillation frequency is not equal to the frequency resolution, but depends on both the frequency resolution and the noise level. \citet{schwarzenberg91} showed that the $1\sigma$ confidence interval of the oscillation frequency is equal to the width of the peak in the power spectrum at the $S-N$ level, where $S$ is the peak height, and $N$ is the average noise level in the vicinity of the peak. To precisely find the peak maximum and its width at the $S-N$ level, the number of frequency components must be significantly increased. This can be achieved by adding zeros at the end of the time series. Then the frequency steps (but not the frequency resolution) are $\Delta f=1/T$, where $T$ is the length of the time series including these zeros. For a combined time series the $S/N$ ratio of the peak may exceed 100. Then the required number of frequency components in the entire power spectrum can be gigantic ($2^{25}$), and the peak itself can contain more than a hundred of frequency components. Using the FFT, an ordinary computer performs the necessary calculations within a minute.

Fig.~\ref{fig2} shows the individual amplitude spectra calculated for the 7 longest individual light curves of J2306 obtained with a time resolution of 8~s.  To better display these amplitude spectra, a significant number of zeros were added at the end of the light curves and the number of frequency components was increased.   As seen, all these amplitude spectra show distinct repeating peaks corresponding to the 464-s oscillation discovered by \citet{halpern18}. These peaks noticeably exceed the surrounding peaks caused by noise, and in some cases they also exceed the noise at the lowest frequencies that is caused by flickering. Thus, although the 464-s oscillation is barely noticeable directly in the light curves, it is easily detected in the amplitude spectra due to repeatability.  As mentioned, a simple and very effective way to distinguish the stable oscillation from intermittent quasi-periodic variations is to check whether the same period is detected in multiple data sets \citep{mukai17}.

\begin{figure}[t]
\includegraphics[width=84mm]{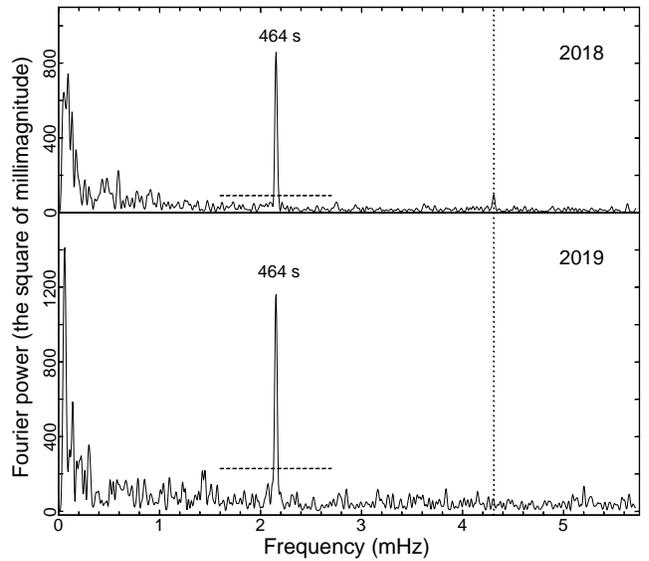}
\caption{Low-frequency parts of the averaged power spectra calculated for two groups of the longest differential light curves of J2306. Two distinct peaks reveal an oscillation with periods of $464.2\pm0.6$ and $464.4\pm0.9$~s in 2018 and 2019, respectively. The vertical dotted line marks the first harmonic of the oscillation.  Two horizontal dashed lines indicate the upper bounds of the 99.95\% confidence interval in the vicinity of the 464-s oscillation.}
\label{fig3}
\end{figure}

Fig.~\ref{fig3} shows two averaged power spectra calculated  by averaging the individual power spectra of the longest individual light curves shown in Fig.~\ref{fig1}. As before, I used a time resolution of 8~s and increased the number of frequency components. Two distinct peaks seen in the power spectra of the 2018 and 2019 data undoubtedly reveal the 464-s oscillation discovered by \citeauthor{halpern18} The statistical significance of detecting this oscillation is very high. These peaks are 9 and 5 times higher than the upper bounds of the 99.95\% confidence intervals for peaks caused by noise in the vicinity of the 464-s oscillation in 2018 and 2019, respectively. The upper bounds were established using the $\chi^2$-distribution \citep[e.g.,][]{bendat86}. In addition, the averaged power spectrum of the 2018 data reveals the first harmonic of the 464-s oscillation. Although the corresponding peak seems small, it nevertheless is 1.7 times higher than the upper bound of the 99.95\% confidence interval for peaks caused by noise in the vicinity of the first harmonic. In the power spectrum of the 2019 data, the first harmonic is not detected. A possible reason is the increased noise level in the power spectrum of the 2019 data compared to the power spectrum of the 2018 data.  Indeed, the average noise levels in the vicinity of the first harmonic are 17 and 42~mmag$^2$ in 2018 and 2019, respectively. Other small peaks visible in the vicinity of the 464-s oscillation are not statistically significant. Peaks visible at frequencies below 0.4~mHz are also not statistically significant due to the increased noise level caused by flickering.

The oscillation periods in the averaged power spectra were $464.2\pm0.6$ and $464.4\pm0.9$~s in 2018 and 2019, respectively. The period errors were determined using the method by \citet{schwarzenberg91}. These errors are overestimated because for averaged power spectra I used the method proposed for individual power spectra. However, in these cases it does not matter much because in the subsequent analysis I determine the oscillation periods with much higher precision. The average semi-amplitudes of the oscillation were $41.8\pm1.5$ and $48.6\pm1.8$~mmag in 2018 and 2019, respectively. The first harmonic detected in the averaged power spectrum of the 2018 data had a period of $232.1\pm0.4$~s. Its semi-amplitude was $14.0\pm0.4$~mmag.

Individual and averaged power spectra make it possible to detect oscillations and determine their amplitudes. However, they cannot be used to precisely measure oscillation periods because these spectra have high noise levels and low frequency resolutions. The power spectra of combined time series consisting of individual light curves and the gaps between them caused by poor weather and seasonal invisibility are more suitable for determining the oscillation period with high precision. Indeed, my longest individual light curve has 4736 points, whereas the total number of points is 49586. Therefore, according to the number of points, the noise level in the power spectrum of the combined time series should be ten times less. The length of the longest individual light curve is 10.5~hr, whereas the length of the combined time series consisting of all individual light curves and gaps is 15 months. Hence, the frequency resolution of the combined time series is 1000 times higher. 

For a coherent oscillation, the power spectrum of the combined time series shows a structure similar to a window function containing different aliases instead of a single peak (e.g.,~\citealt{hellier01}). However, this is more of an advantage than a disadvantage. Indeed, a pair of one-day aliases symmetrically located to the left and right of a certain peak and rising above the surrounding peaks is easily noticeable. This suggests that this peak may be real and that it needs to be analysed for statistical significance. Then the proven statistical significance of both the central peak and its one-day aliases significantly increases the overall statistical significance of the oscillation. This is especially useful for signals detected marginally.  In contrast, the absence of noticeable one-day aliases suggests that the peak is caused by noise or incoherent oscillations. For a signal with a high $S/N$ ratio (more than 100), the power spectrum shows fine details. If these fine details coincide with the fine details of the window function, both in height and frequency, this proves the oscillation coherence during all the observation nights. If the fine details are noticeably different, this suggests that the oscillation is weakly coherent and that the oscillation phase is unstable. Thus, the power spectrum of the combined time series is more informative compared to the power spectrum of an individual time series.

First, I analyzed the 2018 data and the 2019 data included in the two combined time series separately to check the compatibility of the results and then analyzed the overall combined time series consisting of all the data.

\begin{figure}[t]
\includegraphics[width=84mm]{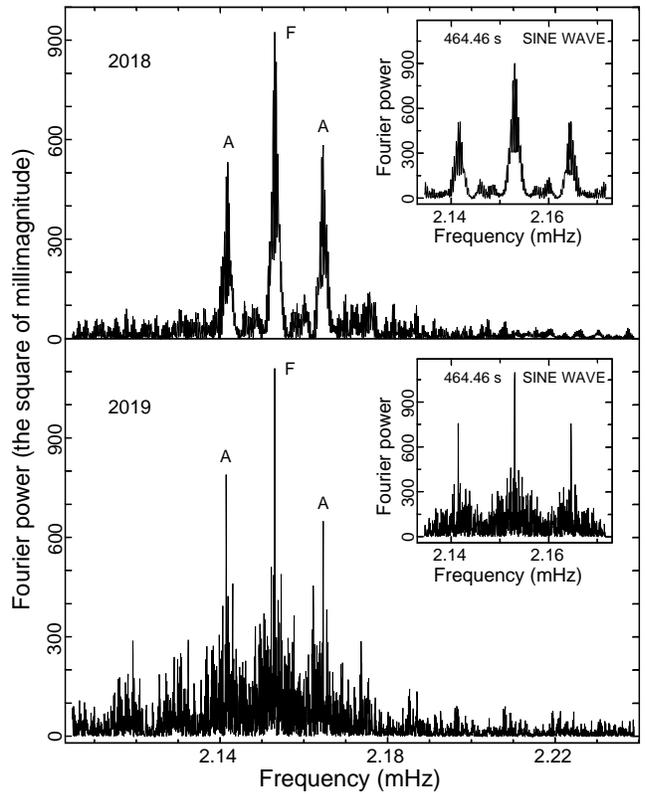}
\caption{Parts of the power spectra calculated for two combined time series consisting of the data of J2306 and gaps according to the observation time. They show an oscillation with periods of 464.4550(20) and 464.4557(9)~s in 2018 and 2019, respectively. The window functions calculated using artificial time series consisting of sine waves and gaps are shown in the insets. The principal peaks are labelled 'F', the one-day aliases are labelled 'A'.}
\label{fig4}
\end{figure}

As mentioned, a time resolution of 8-s seems excessively high to analyze the 464-s oscillation. Therefore two combined time series consisting of the 2018 data and the 2019 data were averaged over 32-s time intervals to diminish the total number of frequency components. In addition, a significant number of zeros were added at the end of these time series to increase the number of frequency components of principal peaks. Each of the two calculated power spectra contained $2^{23}$ frequency components. Fig.~\ref{fig4} shows parts of these power spectra in the vicinity of the 464-s oscillation. The window functions shown in the insets have different shapes due to the peculiarities of the distribution of the light curves over time in 2018 and 2019. Indeed, in the window function of 2018, the principal peak and two one-day aliases consist of several components. In contrast, in the window function of 2019, the corresponding peaks consist of one component, and, in addition, the window function shows many small aliases between them. As seen, the power spectra correspond well to these shapes. The difference in these power spectra highlights that the 464-s oscillation was coherent both during the observations in 2018 and during the observations in 2019.

I determined the noise levels by averaging the power spectra in wide frequency bands before and after the 464-s osculation, excluding the frequencies distorted by aliases. Then I determined the maxima of the principal peaks and their half-widths at the $S-N$ level using Gauss functions fitted to the upper parts of the peaks. According to \citet{schwarzenberg91}, these half-widths are rms errors. 10 and 7 frequency components were above the $S-N$ level in the power spectra of the 2018 and 2019 data, respectively. They showed only a small scatter relative to the Gauss curves. This allowed me to surely measure the half-widths of the peaks at the $S-N$ level. I also used Lorentz functions fitted to the upper parts of the peaks and found no noticeable difference. Thus, I determined the oscillation periods, which were 464.4551(20) and 464.4557(9)~s in 2018 and 2019, respectively. The oscillation semi-amplitudes were 43.1 and 48.3 mmag in 2018 and 2019, respectively.  I also determined the half-widths of the peaks at half-maximum (HWHM), which were 0.023 and 0.009~s in 2018 and 2019, respectively.

\begin{figure}[t]
\includegraphics[width=84mm]{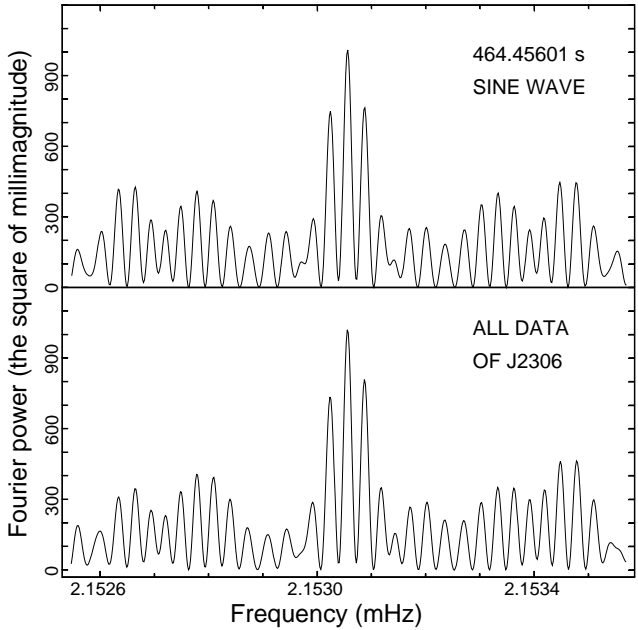}
\caption{Part of the power spectrum calculated for the overall combined time series consisting of all data of J2306 and gaps in the vicinity of the 464-s oscillation. Note the close similarity between the window function shown in the upper frame and this power spectrum. This similarity means that the oscillation was coherent throughout all observations covering 15 months.}
\label{fig5}
\end{figure}

To analyze the overall combined time series, I first used a time resolution of 32~s and calculated $2^{25}$ frequency components in the entire power spectrum. This is the maximum number of frequency components available for calculation. The part of this power spectrum in the vicinity of the 464-s oscillation is shown in the lower frame of Fig.~\ref{fig5}. The main details in this part of the power spectrum are the principal peak and its one-year aliases caused by the large gap between the two observation seasons. One-day aliases go beyond the frequency range shown in Fig.~\ref{fig5}. The fine structure of this power spectrum is very similar to fine structure of the window function shown in the upper frame of Fig.~\ref{fig5}. This means that the 464-s oscillation was coherent throughout all my observations covering 15 months.

The principal peak in the power spectrum of the overall combined time series with a time resolution 32~s had only 2 frequency components above the $S-N$ level. This is not sufficient to surely measure its half-width at the $S-N$ level and to determine the period error. Therefore, I conducted numerical experiments with different time resolutions obtained by averaging the magnitudes in the overall combined time series.  I calculated 12 power spectra using a time resolution in the range 8--96~s. Each power spectrum had the same appearance as in Fig.~\ref{fig5}. The periods in all 12 power spectra were in the range 464.45598--464.45604~s and showed no systematic changes with time resolution. Their average value was 464.45601~s. The average noise levels were close to each other and also showed no systematic changes. The semi-amplitudes of the oscillation in 4 power spectra corresponding to a time resolution of 8--32~s were nearly constant (46.1--46.3 mmag) and systematically decreased from 45.6 to 43.3 mmag due to averaging in 8 power spectra when the time resolution changed from 40 to 96~s. The principal peaks in these 8 power spectra had 3--7 frequency components above the $S-N$ level. Their half-widths at the $S-N$ level determined using Gauss functions fitted to the upper parts of the peaks were close to each other and showed no systematic changes with time resolution. Their average value was 0.00014~s. Thus, the oscillation period determined from the power spectra of the overall combined time series was 464.45601(14)~s. I also determined the HWHM of the principal peaks, which was 0.0018~s. 

Although the principal peak in the power spectra of the overall combined time series is only 25\% higher than its one-year aliases (Fig,~\ref{fig5}), this  does not cause problems with its identification. Indeed, if I consider the nearest one-year alias as the principal peak, then the corresponding period would be incompatible with the period determined using the power spectrum of the combined time series consisting of the 2019 data. Indeed, in this case, the period will be 464.44927(15)~s, and the difference between the two periods will exceed $7\sigma$ (see Table~\ref{periods}).

To verify the results obtained from the power spectra calculated with the FFT, I calculated two power spectra using sine waves fitted to all the light curves folded with the trial frequencies (SWF). By calculating power spectra using the SWF, I can use very small frequency change steps. In addition, such power spectra allow me to determine the error of the oscillation amplitude directly from the fit, whereas a simple method to determine the error of the oscillation amplitude using a single power spectrum calculated with the FFT is unknown.  However, calculating the power spectrum using the SWF is performed 10000 times slower compared to calculating the power spectrum using the FFT. Therefore, power spectra calculated with the SWF can only be obtained for a limited frequency range and cannot be used to accurately measure noise levels by averaging over wide frequency bands. I used a time resolution of 8~s, 60 phase bins and frequency steps of 0.000\,000\,23 and 0.000\,000\,16~mHz. These two power spectra had the same appearance as in Fig.~\ref{fig5}. The oscillation semi-amplitudes in the two power spectra were the same, $47.2\pm2.4$~mmag, and were close to the semi-amplitudes in the power spectra of the overall combined time series calculated with the FFT in cases of a time resolution 8-32~s (46.1--46.3 mmag). Therefore, I used the noise levels determined from the power spectra calculated with the FFT. In two power spectra calculated with the SWF, using Gauss functions fitted to the upper parts of the peaks, I obtained the same oscillation periods and their errors, 464.45600(14)~s. They nearly strictly coincided to those in the power spectra of the overall combined time series calculated with the FFT. This demonstrates that, using a gigantic number of frequency components in the power spectra ($2^{25}$), I did not make gross errors. 

\begin{table}[t]
{\small
\caption{The oscillation periods determined using the power spectra of the combined time series.}
\label{periods}
\begin{tabular}{@{}l l l l l l l }
\hline
\noalign{\smallskip}
Time & S/N  & Period & HWHM & Error & Dev. \\
span &          & (s)       & (s)         & (s)     &    \\
\hline
2018 &  170  & 464.4550    & 0.0230   &  0.0020   & $0.5\sigma$    \\
2019 &  130  & 464.4557    & 0.0088   &  0.0009    & $0.3\sigma$    \\
Total &  270  & 464.45601  & 0.00180  &  0.00014  & -- \\
\hline
\end{tabular} }
\end{table}

Table~\ref{periods} shows the oscillation periods determined using the power spectra of the combined time series  consisting of the 2018 data, the 2019 data and all data covering 15 months. As seen, the precision of the periods increases with the increase of the $S/N$ ratio and with the improvement of the frequency resolution, which is approximately equal to the HWHM of the peaks. Within the errors, all periods are compatible with each other.

\begin{figure}[t]
\includegraphics[width=84mm]{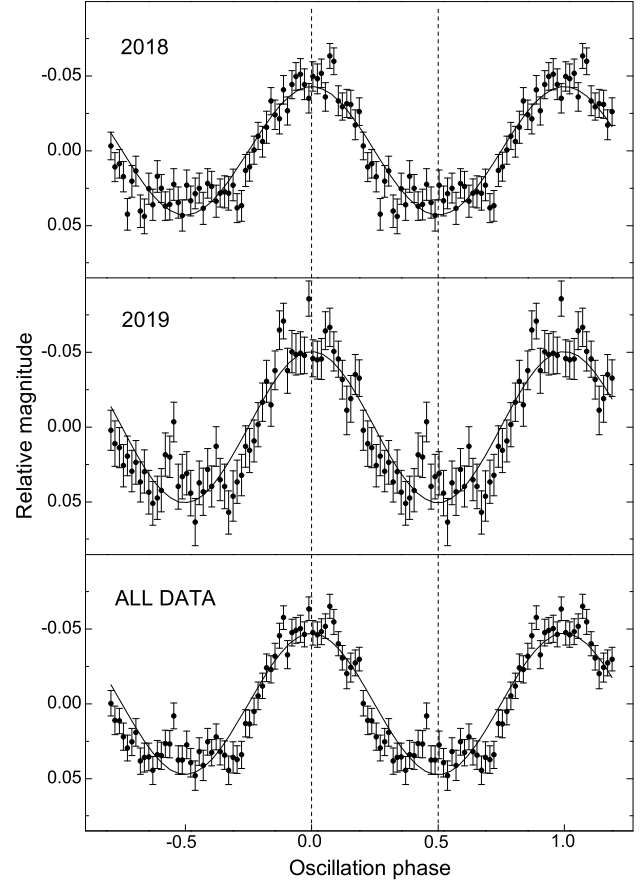}
\caption{Oscillation pulse profiles obtained by folding the light curves of J2306 with a period of 464.45601~s. All profiles are symmetrical and show noticeably wider minima compared to maxima.}
\label{fig6}
\end{figure}

Fig.~\ref{fig6} shows the pulse profiles of the 464-s oscillation obtained by folding the light curves with a period of 464.45601~s. I used a time resolution of 8~s and 60 phase bins. Thus, each phase bin was 7.74~s and was close to the time resolution. To demonstrate the pulse profile stability, I folded the light curves obtained in 2018, in 2019 and using all data.  As seen, all profiles are symmetrical and show noticeably wider minima compared to maxima. Indeed, in all folded light curves, most consecutive points to the left and right of the minima are below the sine wave by 1--2$\sigma$, whereas most consecutive points near the minima are above the sine wave by 1--2$\sigma$. In contrast, the points around the maxima are distributed quite uniformly relative to the sine wave. This noticeable difference from the sine wave is the reason for the first harmonic of the 464-s oscillation in the power spectra of the 2018 data. Although the first harmonic was not detected in the power spectra of the 2019 data, this can be caused by the larger noise level in 2019 and cannot be caused by a pulse profile change. Indeed, as seen in Fig.~\ref{fig6}, all pulse profiles are similar to each other. 

The oscillation semi-amplitudes determined using sine waves fitted to the pulse profiles were $43.0\pm2.3$ and $50.6\pm3.2$~mmag in 2018 and 2019, respectively. Within the errors, they are compatible with the semi-amplitudes in the averaged power spectra (Fig.~\ref{fig3}). The oscillation semi-amplitude determined using all data was $47.2\pm2.4$~mmag and obviously coincided with the semi-amplitudes found in the power spectra calculated using sine waves fitted to the light curves folded with the trial frequencies (see above).

The high precision of the oscillation period allowed me to derive an oscillation ephemeris with a long validity. Because individual light curves contain significant noise, I determined the maximum time using all the light curves folded with a period of 464.45601~s and referred it to the middle of the observations. Because the folded light curve is symmetrical (Fig.~\ref{fig6}), I used a sine wave fitted to the folded light curve to precisely determine the maximum time. So I obtained the following tentative ephemeris:

{\scriptsize
\begin{equation}
BJD_{\rm TDB}(max)= 245\,8624.794\,592(44) + 0.005\,375\,6483(16) E.
\label{ephemeris1}
\end{equation} }

\begin{figure}[t]
\includegraphics[width=84mm]{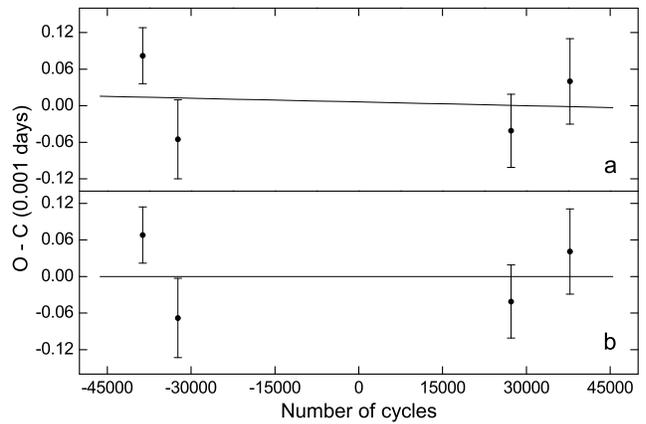}
\caption{(a) (O--C) diagram calculated for the tentative ephemeris obtained using the period determined from the power spectrum of the overall combined time series and the maximum time determined from all the light curves of J2306 folded with this period. This (O--C) diagram shows the slope and offset along the ordinate. (b) (O--C) diagram calculated for the ephemeris corrected using the coefficients of a first-order polynomial fitted to the (O--C) values.}
\label{fig7}
\end{figure}

To verify ephemeris~\ref{ephemeris1}, I divided all the data into four groups (see Table~\ref{groups}) and folded the light curves in each group with a period of 464.45601~s. These folded light curves looked similar to light curves shown in Fig.~\ref{fig6}. Using sine waves fitted to the folded light curves, I determined the maximum time and semi-amplitude in each group and referred them to the middle of the corresponding observations. In accordance with ephemeris~\ref{ephemeris1}, I calculated the (O--C) values and the numbers of the oscillation cycles (Table~\ref{groups}). The (O--C) diagram is shown in Fig.~\ref{fig7}a. This (O--C) diagram reveals the slope and offset along the ordinate and obeys the linear relation: $O-C=0.000\,006(40) - 0.000\,000\,0002(12) E$. The coefficients in this relation are much less than their rms errors. Nevertheless, I corrected ephemeris~\ref{ephemeris1}  using these coefficients because the first-order polynomial fitted to the (O--C) values allows me to independently determine the oscillation period and its error. So I obtained the final corrected ephemeris:

{\scriptsize 
\begin{equation}
BJD_{\rm TDB}(max)= 245\,8624.794\,598(44) + 0.005\,375\,6481(12) E.
\label{ephemeris2}
\end{equation} }


\begin{table*}[t]
{\small
\caption{Verification of the tentative ephemeris of the 464-s oscillation using the data divided into four groups.} 
\label{groups}
\begin{tabular}{@{}l l l l c}
\hline
\noalign{\smallskip}
Time         & BJD$_{\rm TDB}$(max)  &   Semi-amp.     & N. of    & O--C$\times10^{3}$   \\
span         & (-2458000)                     &    (mmag)       & cycles   &   (days)    \\
\hline
2018 Oct.9--Nov.9        & 416.961359(46)  &  $42.7\pm2.3$   & $-38662$  &  $+0.082(46)$  \\
2018 Nov.10--Dec.15   & 450.806304(65)  &  $43.4\pm3.2$    & $-32366$ &  $-0.055(65)$  \\
2019 Sep.8--Nov.19       & 771.270216(60)  &  $51.8\pm3.6$    & $+27248$  & $-0.041(60)$  \\ 
2019 Nov.20--Dec.30   & 828.123153(70)  &  $49.4\pm4.0$    & $+37824$   & $+0.040(70)$  \\ 
\hline 
\end{tabular} }
\end{table*}

The (O--C) diagram obtained according to \\ ephemeris~\ref{ephemeris2}  is shown in Fig.~\ref{fig7}b. As seen, the slope and offset along the ordinate are completely excluded. The period determined using the coefficients of a first-order polynomial fitted to the (O--C) values is 464.45600(10)~s. This period nearly strictly coincides with the period determined using the power spectra of the overall combined time series (Table~\ref{periods}). Its error turned out to be somewhat less than the error determined from the power spectra of the overall combined time series using the method by \citet{schwarzenberg91}. 


A first-order polynomial fitted to the (O--C) values must correct any false period if this period does not cause ambiguity in the cycle count. To demonstrate this, I performed numerical experiments. I constructed three ephemerides using false periods, which were 464.4553, 464.4551 and 464.4542~s, and using the data divided into four groups and folded with these false periods. The corresponding (O--C) diagrams showed significant slopes. I excluded these slopes using the coefficients of a first-order polynomial fitted to the (O--C) values and obtained corrected periods, which were 464.45599(12), 464.45598(12) and 464.45596(15)~s, respectively. It seems that a larger slope gives a larger deviation from the period determined using the power spectra of the overall combined time series and a larger period error. Nevertheless, in all three cases, the periods and their errors determined using a first-order polynomial fitted to the (O--C) values turned out to be close to the periods and their errors determined using the power spectra of the overall combined time series. These numerical experiments prove that two independent methods (i.e., the power spectrum and the first-order polynomial fitted to the (O--C) values) give periods that coincide within their errors. Therefore, these periods and their errors are correct.

I determined the noise levels by averaging the power spectra over a wide frequency bands before and after the 464-s oscillation. Because the noise decreases non-linearly with frequency due to flickering, the noise level near the 464-s oscillation may be somewhat overestimated, and therefore the period error determined using the method by \citet{schwarzenberg91} may also be somewhat overestimated. Therefore, finally I accept the period and its error determined by the coefficients of a first-order polynomial fitted to the (O--C) values, which were calculated using ephemeris~\ref{ephemeris1}, 464.45600(10)~s.


The formal validity of an ephemeris is the time during which the accumulated error of the period reaches one oscillation cycle. Obviously, the formal validity means the time span during which the ephemeris can be used without ambiguity in the cycle count. According to the period error, the formal validity of ephemeris~\ref{ephemeris2}  is 70 years (a $1\sigma$ confidence level).


\begin{figure}[t]
\includegraphics[width=84mm]{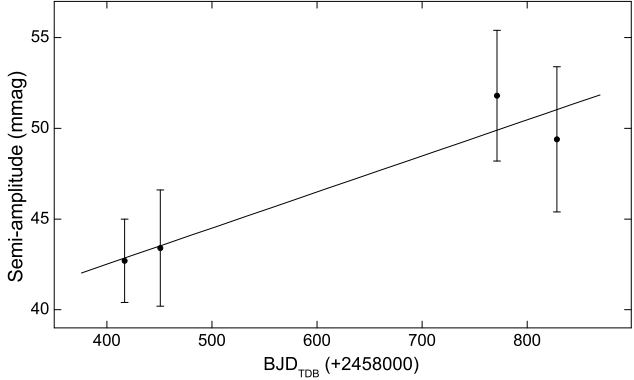}
\caption{Semi-amplitudes of the 464-s oscillation as a function of time. They were determined using the data of J2306 divided into four groups and folded with a period of 464.45601~s.  A first-order polynomial fitted to these semi-amplitudes reveals a statistically significant slope.}
\label{fig8}
\end{figure}

Although the peaks corresponding to the 464-s oscillation showed noticeably different heights in the averaged power spectra (Fig.~\ref{fig3}), this does not prove that the amplitude of the oscillation varies between 2018 and 2019. Indeed, the difference of semi-amplitudes in these power spectra is $6.8\pm2.3$~mmag and reaches only $3\sigma$. However, to calculate the averaged power spectra, I used only longest light curves, but not all the data. The light curves folded with the oscillation period (Fig.~\ref{fig6}) show different amplitudes in 2018 and 2019. It also does not prove that the amplitude varies between 2018 and 2019 because the difference of semi-amplitudes is $7.6\pm3.9$~mmag and reaches only $1.9\sigma$. However, as seen in Fig.~\ref{fig6}, the oscillation profiles are not strictly sinusoidal and hence the errors determined by sine waves fitted to the profiles may be overestimated. To prove the reality of the amplitude change of the 464-s oscillation between 2018 and 2019, I considered the semi-amplitudes determined using the data divided into four groups (see Table~\ref{groups}).  Fig.~\ref{fig8} shows these semi-amplitudes as a function of time. A first-order polynomial fitted to these semi-amplitudes reveals a slope of $0.0199\pm0.0048$ mmag d$^{-1}$. This slope is statistically significant because it is 4 times larger than its rms error. Thus, the oscillation amplitude change between 2018 and 2019 is real.

Using two combined time series containing the 2018 data and the 2019 data, I attempted to measure the period of the first harmonic of the 464-s oscillation. Near the first harmonic, the power spectrum of the 2018 data showed the principal peak and two one-day aliases, which were multicomponent  in accordance with the window function (see the upper inset in Fig.~\ref{fig4}). This indicated a coherent oscillation. However, due to the high noise level, it was impossible to identify the principal component and to precisely measure the period. Near the first harmonic in the power spectrum of the 2019 data, I found a peak that strictly coincides with the first harmonic. However, many peaks caused by noise in the vicinity of this peak had similar heights. So I could not identify the one-day aliases. As seen in the insets in Fig.~\ref{fig4}, one-day aliases should be 30--40\% less than a principal peak. Obviously, to detect one-day aliases, the principal peak must be at least 2 times higher than the surrounding noise peaks. In 2018, the maximum height of peaks caused by noise in the vicinity of the first harmonic was 47~mmag$^2$, whereas the height of the principal peak was 81~mmag$^2$ and was roughly two times higher. In contrast, in 2019, the maximum height of peaks caused by noise in the vicinity of the first harmonic was 88~mmag$^2$. Therefore, in 2019, both the principal peak and the one-day aliases corresponding to the first harmonic having the same amplitude as in 2018 could not be detected due to the higher noise level.

At low frequencies, the averaged power spectra did not reveal statistically significant peaks that could indicate the orbital period (Fig.~\ref{fig3}). The power spectra of the two combined time series containing the 2018 data and the 2019 data, did not show one-day aliases that could indicate a coherent oscillation corresponding to the orbital period. The maximum heights of peaks caused by noise in these power spectra at low frequencies were 512 and 2420~mmag$^2$ in 2018 and 2019, respectively. As mentioned, to detect one-day aliases, the principal peak must be at least two times higher than the maximum height of the peaks caused by noise. Thus, the thresholds for detecting one-day aliases caused by orbital variability were 1024 and 4840~mmag$^2$ in 2018 and 2019, respectively. These thresholds correspond to semi-amplitudes of 45 and 98~mmag in 2018 and 2019, respectively.

A distinctive feature of IPs is the presence of orbital sidebands. The averaged power spectra (Fig.~\ref{fig3}) did not reveal statistically significant peaks that could be attributed to additional oscillations with periods close to the period of 464~s. To detect additional oscillations due to their coherence indicated by one-day aliases, I analyzed two combined time series consisting of the 2018 data and the 2019 data. Additional oscillations may have relatively small amplitudes and may be hidden in the fine structure of the window function of the 464-s oscillation. Therefore, I used the well-known method of subtracting the main oscillation from the data. In accordance with the observations in 2018 and 2019, I constructed two artificial combined time series consisting of a sine wave and gaps. These two artificial time series folded with a period of 464.45601~s showed the same amplitudes and phases as the real combined time series folded with this period. Then I subtracted these artificial time series from the real time series and obtained two time series that were pre-whitened for the 464-s oscillation.

\begin{figure}[t]
\includegraphics[width=84mm]{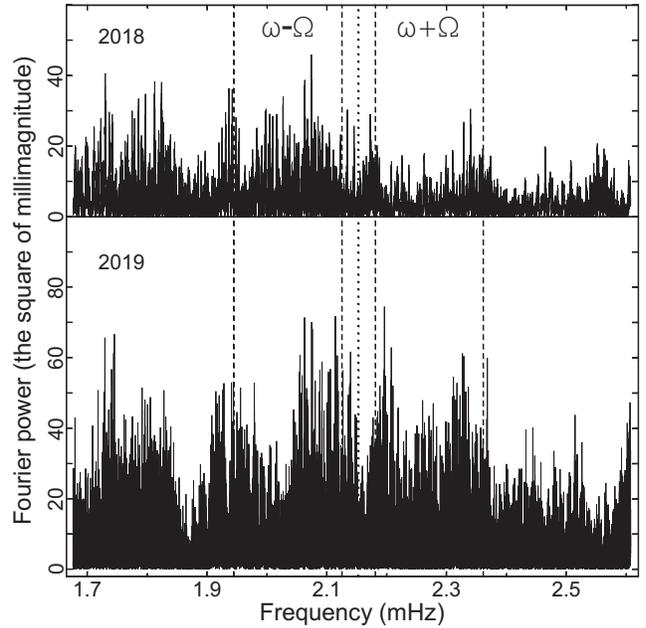}
\caption{Power spectra of two combined time series consisting of the data of J2306, which were pre-whitened for the 464-s oscillation. The dotted line marks the frequency of the 464-s oscillation. Dashed lines indicate possible frequency bands for sideband oscillations assuming that the 464-s period is the spin period of the white dwarf and that the orbital period is in the range 80~min--10~hr.}

\label{fig9}
\end{figure}

The parts of two power spectra of the pre-whitened combined time series are shown in Fig.~\ref{fig9}. As seen, the 464-s oscillation is completely excluded. This, however, did not allow me to detect additional oscillations. Indeed, these parts of the power spectra only show peaks caused by noise without any hints of one-day aliases. The maximum heights of these peaks are 46 and 74~mmag$^2$ in 2018 and 2019, respectively. As mentioned, to detect one-day aliases, the principal peak must be at least two times higher than the maximum height of peaks caused by noise. Hence, the thresholds for detecting one-day aliases of additional oscillations were 92 and 148~mmag$^2$ in 2018 and 2019, respectively. Thus, the power spectra of the combined time series did not reveal additional oscillations with periods close to the period of 464~s, the semi-amplitudes of which exceed 14 and 17 mmag in 2018 and 2019, respectively.   In addition, analyzing the power spectra up to the Nyquist limit (16~s), I did not find any other coherent oscillations in addition to the 464-s oscillation and its first harmonic.

\section{Discussion}

I conducted extensive photometric observations of J2306 over 22 nights covering 15 months. The total duration of observations was 110~hr. I clearly detected an oscillation with a period of 464.45600(10)~s. Within the errors, the period of this oscillation coincides with the period of the oscillation discovered in J2306 by \citet{halpern18}, which was 464.452(4)~s. The period precision obtained in my observations is 40 times higher than the period precision obtained by \citeauthor{halpern18} This increase in precision is achieved by increasing the total duration and coverage of observations. As seen in Fig.17 in \citeauthor{halpern18}, the total duration and coverage of their observations are 12~hr and 21~d. The total duration of my observations is 9 times longer and the coverage of my observations is 22 times longer. The high precision of the oscillation period is the most important result of my observations of J2306. The detected oscillation was coherent throughout all my observations covering 15 months. This is the second important result of my observations of J2306.

Short-period X-ray oscillations that could confidently classify J2306 as an IP have not yet been detected. As noted by \citet{kuulkers06}, in such cases, the presence of orbital sidebands in addition to a stable optical short-period oscillation can be a strong argument for inclusion in the IP list. Unfortunately, despite long-term observations, I could not detect additional oscillations in the power spectra of J2306 that could be identified with orbital sidebands.  \citeauthor{halpern18} also found no orbital sidebands. Nevertheless, on the basis of optical observations performed by \citeauthor{halpern18}, J2306 was reckoned among the confirmed IPs in the IP Homepage by K. Mukai (https://asd.gsfc.nasa.gov/Koji.Mukai/iphome/ iphome.html).

I examined the IP Homepage and references therein to identify IPs in which X-ray short-period oscillations were not detected. The total number of the ironclad and confirmed IPs in the IP homepage is 71. In addition to J2306, 6 of them did not show both X-ray short-period oscillations and additional optical oscillations that could be identified with orbital sidebands. Hence, an optical short-period oscillation with a stable period is considered sufficient to recognize a CV as an IP, and the presence of orbital sidebands is not necessary. However, the very high stability of the optical  periods were proved for only three of these 6 IPs due to photometric observations during many observation nights covering several observing seasons. These IPs are DQ Her with $P_{\rm spin}=71.065\,5828(3)$~s \citep{zhang95}, V455~And  with $P_{\rm spin}= 67.619\,703\,96(72) $~s \citep{mukadam16} and V1460~Her  with $P_{\rm spin}=38.871\,256\,78(10)$~s \citep{pelisoli21}. The optical periods of other three IPs were measured with much lower precision due to short observations covering only a few nights or even a single night. These IPs are LAMOST~J024048.51+195226.9 with $P_{\rm spin}= 24.9328(38)$~s \citep{pelisoli22}, CXOGBS~J174954.5-294335 with $P_{\rm spin}= 503.32(3)$~s \citep{johnson17} and \\ PBC~J1841.1+0138 with $P_{\rm spin}=311.805(4) $~s \\ \citep{halpern22}. 

However, insufficiently long photometric observations, which prove the oscillation stability for only a few nights, can lead to an incorrect interpretation of the period. This is demonstrated by the observations of V455~And. Using observations of V455~And, most of which consisted of 7 consecutive nights, \citet{araujo05} detected an oscillation with a period of 67.2~s. This oscillation seemed coherent, and the 67.2~s period was attributed to the spin period of the white dwarf. However, using extensive photometric observations obtained with 9 telescopes, \citet{gansicke07} found that the 67.2-s oscillation was incoherent, and another oscillation with a period of 67.62~s, which had a much less amplitude, was indeed coherent. I also observed V455 And and found that the oscillation with a period of 67.2~s was weakly coherent and could imitate a coherent oscillation in the power spectrum of a few consecutive nights \citep{kozhevnikov15}.

My observations of J2306 reveal the coherence of the 464-s oscillation over 15 months. This is obvious due to the close similarity between the power spectrum of the overall combined time series and the corresponding window function, which are calculated for a gigantic number of frequency components ($2^{25}$). Because each peak in Fig.~\ref{fig5} consists of about 100 frequency components, all the details of the power spectrum and the window function are clearly visible. The close similarity between the two means that, like the sine wave used to calculate the window function, the 464-s oscillation was coherent throughout all my observations covering 15 months. The coherence of the 464-s oscillation throughout all my observations also follows from Fig.~\ref{fig9}, which shows that this oscillation can be eliminated by subtracting a sine wave having the appropriate amplitude and phase (e.g., \citealt{gansicke07}). The high degree of coherence of the 464-s oscillation greatly increases the probability that J2306 is an IP.

In most IPs, the period of the X-ray oscillation with the largest amplitude is attributed to the spin period of the white dwarf because the X-ray orbital sidebands normally have lesser amplitudes (e.g., \citealt{demartino20}). In contrast, the amplitudes of optical spin oscillations in the known IPs can be both larger and less than the amplitudes of optical sideband oscillations \citep{warner95}. Because the short-period X-ray oscillations were not detected in J2306, it is difficult to definitely conclude whether the 464-s period is the spin period of the white dwarf or it is an orbital sideband. However, the pulse profiles of sideband oscillations often show significant variability, and changes of the accretion disc structure are considered plausible reasons for such variability \citep{woerd84}. Indeed, I observed significant changes of the sideband pulse profile in MU Cam from year to year, whereas the spin pulse profile remained unchanged (see Fig.~8  in \citealt{kozhevnikov16}). Thus, because the pulse profile of the 464-s oscillation showed no noticeable changes during 2018 and 2019 (Fig.~\ref{fig6}), the period of 464.45600(10)~s is most probably the spin period of the white dwarf.


Some IPs show optical short-period oscillations with very low amplitudes. I observed V709~Cas and EI~UMa, in which the semi-amplitudes of the oscillations were about 5 and 8 mmag, respectively, and these oscillations could be surely detected only using a combined time series consisting of several observation nights \citep{kozhevnikov01, kozhevnikov10}. The oscillation periods and phases in such IPs are difficult to measure with high precision due to noise.  In contrast, the semi-amplitude of the 464-s oscillation in J2306 was about 10 times larger, and this oscillation was easy to detect even in one observation night. Due to the relatively low noise level, the oscillation period can be measured with high precision. In addition, the pulse profile of the 464-s oscillation is symmetrical (Fig.~\ref{fig6}), and the oscillation phase can be determined with high precision by fitting a sine wave. Thus, J2306 is a good candidate for future studies of changes of the oscillation period.

Long-term tracking of the oscillation period in J2306 can be performed by direct measurements of the period. If I assume that the white dwarf in J2306 has d$P$/d$t=10^{-11}$, which seems typical of slowly spinning IPs (see Table~1 in \citealt{warner96}), then the period change will be 0.0003~s per year and will be much less than the period error obtained from observations during one observing season (0.001--0.002~s, Table~\ref{periods}). Obviously, to achieve much higher precision, photometry must be carried out over two seasons separated by one year, and each season should cover about 10 nights to eliminate the problem with aliases. Then observations over 6 years will give 3 precise period measurements that could reliably detect a spin period change. 

Instead of direct period measurements, long-term tracking of the oscillation period in J2306 can be performed by analyzing the (O--C) values obtained using ephemeris~\ref{ephemeris2}.  To estimate the possible (O--C) values, I used the following formula \citep{breger98}:
\begin{equation}
{\rm (O - C)} = 0.5 \, \frac{1}{P} \, \frac{{\rm d}P}{{\rm d}t} \, t^2.
\label{breger}	
\end{equation}
Using formula~\ref{breger} and d$P$/d$t=10^{-11}$, I calculated (O--C) = 11~s for one year. The (O--C) error obtained using several observation nights can be two times less (Table~\ref{groups}). Hence, the (O--C) value can reveal a noticeable period change even within one year. Of course, monotonic period changes can only be detected due to non-linear changes of (O--C) values over time (e.g., \citealt{breger98}). Therefore, the observations should be carried out for several years. However, because a sufficiently precise (O--C) value can be obtained using only a few long observation nights, the analysis of the (O--C) values seems less laborious compared to direct measurements of the period.

\section{Conclusions}

I performed extensive photometric observations of J2306 over 22 nights in 2018 and 2019. The total duration of the observations was 110~hr. I obtained the following results:
\begin{enumerate}
\item I clearly detected an oscillation with a period of 464~s that was recently discovered in J2306 by \citet{halpern18}. This oscillation was revealed in each long observation night.  
\item The oscillation was coherent throughout all my observations covering 15 months. This greatly increases the probability that J2306 is an intermediate polar.
\item The long coverage of observations allowed me to precisely determine the oscillation period, which was $464.45600\pm0.00010$~s. 
\item The oscillation semi-amplitude was large and showed changes from $41.8\pm1.5$~mmag in 2018 to $48.6\pm1.8$~mmag in 2019.

\item The oscillation pulse profile was symmetrical with a noticeably wider minimum compared to the maximum and showed no noticeable changes during 2018 and 2019. 

\item  I derived the oscillation ephemeris with a formal validity of 70~yr. This ephemeris can be used for future studies of oscillation period changes.

\end{enumerate}

\section*{Acknowledgments}

The work was supported by the Ministry of Science and Higher Education of the Russian Federation, FEUZ-2020-0030. This research has made use of NASA's Astrophysics Data System Bibliographic Services.

\section*{Data availability} 
The datasets generated during and/or analysed during the current study are available from the corresponding author on reasonable request.

\vspace{1 cm}

This version of the article has been accepted for publication, after peer review but is not the Version of Record and does not reflect post-acceptance improvements, or any corrections. The Version of Record is available online at: https://doi.org/10.1007/s10509-022-04159-x. Use of this Accepted Version is subject to the publisher’s Accepted Manuscript terms of use https://www.springernature.com/ \\ gp/open-research/policies/accepted-manuscript-terms.

\end{document}